\documentclass[journal,comsoc]{IEEEtran}
\usepackage[T1]{fontenc}

\usepackage{amssymb,amsmath,amsfonts,amsthm}
\usepackage{mathrsfs}
\usepackage{cite}
\usepackage{array}
\usepackage{tabularx}
\usepackage[table]{xcolor}
\usepackage{color}
\usepackage{mathtools}
\usepackage{caption}
\usepackage{subcaption}
\usepackage{mathtools}
\usepackage{tikz,pgf}
\usetikzlibrary{shapes.geometric}
\usetikzlibrary{quotes}
\usepackage{graphicx}
\usepackage{bbm}
\usepackage[utf8]{inputenc}
\usepackage{algorithm}
\usepackage{algcompatible}
\usepackage{xcolor}
\definecolor{darkblue}{rgb}{0,0.08,0.45}
\usepackage{hyperref}
\hypersetup{
	colorlinks=true,
	citecolor=darkblue,
}

\newcommand{\minimize}{\mathop{\rm minimize}\limits}

\usepackage[colorinlistoftodos,prependcaption,backgroundcolor=black!5!white,bordercolor=red]{todonotes}
\usepackage{marginnote}

\newcommand{\mytodo}[2][]{{%
		\let\marginpar\marginnote
		\reversemarginpar
		\renewcommand{\baselinestretch}{0.8}%
		\todo[#1]{#2}}}

\IEEEoverridecommandlockouts
\ifCLASSINFOpdf
\else
\fi
\usepackage[cmintegrals]{newtxmath}
\begin{document}
	\title{Over-the-Air Computation with Multiple Receivers: A Space-Time Approach}
	
	\author{Zheng Chen and Yura Malitsky
		\thanks{Z. Chen is with the Department of Electrical
			Engineering,  Link\"{o}ping University, 58183 Linköping,
			Sweden (email: zheng.chen@liu.se). \par Y. Malitsky is with the Department of
			Mathematics, University of Vienna, 1090 Wien, Austria (email: 
		yurii.malitskyi@univie.ac.at). \par This work was supported in part by
			Zenith, ELLIIT, Swedish Research Council (VR), and by the Wallenberg Al, Autonomous
			Systems and Software Program (WASP) funded by the Knut and
			Alice Wallenberg Foundation, no 305286. The work of Y. Malitsky was performed while he was with Link\"{o}ping University.}}
	
	\maketitle
	
	\begin{abstract}
		The emerging concept of Over-the-Air (OtA) computation has shown great potential for achieving resource-efficient data aggregation across large wireless networks. However, current research in this area has been limited to the standard many-to-one topology, where multiple nodes transmit data to a single receiver. In this study, we address the problem of applying OtA computation to scenarios with multiple receivers, and propose a novel communication design that exploits joint precoding and decoding over multiple time slots. To determine the optimal precoding and decoding vectors, we formulate an optimization problem that aims to minimize the mean squared error of the desired computations while satisfying the unbiasedness condition and power constraints. Our proposed multi-slot design is shown to be effective in saving communication resources (e.g., time slots) and achieving smaller estimation errors compared to the baseline approach of separating different receivers over time. 

	\end{abstract}

	\section{Introduction}	
	Over-the-Air (OtA) computation has recently emerged as a promising solution for efficient data aggregation over distributed nodes \cite{ota-iot, chen2022over}. It exploits the superposition property of analog signals in wireless channels, without the need of encoding continuous-valued information data into discrete-valued digital symbols as in traditional digital communication systems. 
	This concept is particularly relevant when the goal of communication is to aggregate multiple data streams such that the message is approximately received, instead of receiving each data stream correctly without errors.
	
	The fundamental theory behind OtA computation can be traced back to the distributed computation of nomographic functions \cite{buck1982nomographic}. Previously, this concept has been mainly investigated for data aggregation over wireless sensor networks \cite{nomographic_function} and the joint source-channel design for distributed computation of functions over multiple access channels \cite{computation-mac}. More recently, it has been considered for statistical estimation \cite{ota-estimation}, federated learning \cite{yang2020federated,  zhu2019broadband, sery2021over, rate-optimization}, wireless control \cite{ota-control}, and many other applications. The optimal transmitting-receiving scaling design under peak power constraints is investigated in \cite{ota-opt-scaling}, where the scaling laws of the computation error and the power consumption are also provided. Shifting from the standard single-channel OtA model with independent data, several recent works have considered extension to cases with broadband channels \cite{ota-broadband}, with correlated signals \cite{ota-correlated}, and with reconfigurable intelligent surface \cite{star-ris}. 
	
	\vspace{-0.1cm}
	\subsection{OtA Computation over Multiple Access Fading Channels}
	The principle of a classical OtA computation system with multiple senders and one receiver (fusion center) can be described as follows. Assume there are $N$ senders in the system, and the goal is to compute the arithmetic mean of the data from all senders, i.e.,  
	$\theta=\frac{1}{N}\sum_{i=1}^{N}s_i$, 
	where $s_i$ represents the data symbol from the $i$-th sender.
	In general, the objective of computation can be any nomographic function of the data samples from all sender. We consider the arithmetic mean only for the purpose of illustrating the basic concept.
	
	Let $h_i$ denote the channel gain from the $i$-th sender to the receiver. To align the received signal phase and amplitude at the receiver, each sender applies a scaling factor (precoder) $b_i=\frac{\eta}{N\cdot h_i}$,
	where $\eta$ is an amplitude scaling factor. Under a given transmit power constraint $|b_i s_i|^{2}\leq P_{\max}, \forall i$, we have
	\begin{equation}
		\eta=\sqrt{P_{\max}}\min\limits_{i=1,\ldots, N} \left\{\frac{N|h_i|}{|s_i|}\right\}.
		\label{eq:eta}
	\end{equation}
	
	Then, the precoded signal is directly transmitted over the channel using analog modulation and the received superimposed signal at the receiver is given as
	\begin{equation}
		y=\sum_{i=1}^{N}h_i b_i s_i +n,
		\label{eq:channel-single}
	\end{equation}
	where $n$ is the additive white Gaussian (AWGN) noise with variance $\sigma^2$.
	The receiver decodes the signal and computes the estimated aggregated data as
	$\hat{\theta}=y/\eta$.    
	
	The performance of the computation is usually evaluated by the mean squared error (MSE) of the aggregated data
	\begin{equation}
		\text{MSE}=\mathbb{E}[|\hat{\theta}-\theta|^2].
	\end{equation}
	With perfect channel state information at the sender side, the MSE (or the effective noise variance) of the computation is equal to $\sigma^2/\eta^2$.

		\vspace{-0.1cm}
	\subsection{Related Works} 
	The vast majority of existing research on OtA computation considers the standard topology with multiple senders and one receiver, where the communication/computation is done within one slot. Multi-slot OtA computation (with one receiver) was investigated in \cite{tang2020multi}, as a method to exploit channel diversity over time. The problem gets more interesting and challenging when there are multiple spatially distributed receivers in the system, while each receiver needs to compute some functions of the data samples from its associated senders. The precoding design is less straightforward since one sender cannot adapt its precoder to multiple channels with one-shot computation. 
	The idea of using multi-slot OtA computation in fully decentralized stochastic gradient descent systems with multiple receivers was explored in \cite{ota-sgd}, while the analysis was restricted to the case with one active receiver at a time. This consideration is equivalent to dividing the whole network into many parallel sub-networks where each sub-network follows the standard many-to-one topology. 
	
		\vspace{-0.1cm}
	\subsection{Contributions}
	In this work, we propose a \emph{space-time} approach for OtA computation with multiple senders and multiple receivers communicating over multiple time slots to achieve desired computation at each receiver. A main motivation for our proposed approach is that the multi-slot joint precoding and decoding design exploits the degrees of freedom both in space (different nodes) and in time (different slots). 
	The effectiveness of our proposed design is verified through simulations, which show notable gain in reducing communication time and estimation error as compared to the baseline approach of separating different receivers over time.

	\textit{Notation:} We use $|\mathcal{S}|$ for the cardinality of a set $\mathcal{S}$, $\|\mathbf{x}\|$ for the norm of a vector $\mathbf{x}$, $\mathbf{x}^{*}$ for the conjugate of a complex vector $\mathbf{x}$, $\mathbf{x}^\top$ for the transpose of $\mathbf{x}$,   $\mathbf{x}^H=(\mathbf{x}^{*})^\top$ for the conjugate transpose of $\mathbf{x}$, $\mathbf{a}^{H}\mathbf{b}=\sum_{i}a_i^{*} b_i$ for the dot product of two complex vectors $\mathbf{a}$ and $\mathbf{b}$. 
	
	\section{System Model}
	We consider an OtA computation system where multiple senders communicate simultaneously with multiple receivers over multiple access fading channels. The goal of communication is to compute at each receiver the arithmetic mean of the data samples from the set of senders that are connected to this receiver. 
	
\begin{figure}[h]
		\centering
		\tikzset{unode/.style = {
    circle, 
    draw=cyan!30!black, 
    thick,
    fill=cyan!80!black,
    inner sep=3.0pt,
    minimum size=2.3pt } }

\tikzset{uedge/.style = {
    <-,
    draw=cyan!20!black, 
    very thick} }

\tikzset{triangle/.style = {
  regular polygon,
  regular polygon sides=3,
    draw=cyan!30!black,
    thick,
    fill=cyan!60!green,
    inner sep=2.3pt,
   minimum size=2.3pt
    }
  }

\begin{tikzpicture}[scale=2.5, >=stealth, every edge quotes/.append
  style = {auto, font=\footnotesize, inner sep=0pt}] 
  \begin{scope}
    \newcommand\m{3}    
    \newcommand\n{2}    

    \foreach \i in {1,...,\m}{
      \node[unode] (t\i) at (-0.5*\m + \i,1) {};
    }
    \foreach \j in {1,...,\n} {
      \node[triangle] (b\j) at (-0.5*\n + \j,0) {};
    }
    \path[uedge] (b1) edge[pos=0.9,"$h_{11}$"] (t1) ;
    \path[uedge] (b2) edge[pos=0.7,"$h_{12}$"] (t1) ;
    \path[uedge] (b1) edge[pos=0.9,"$h_{21}$"] (t2) ;
    \path[uedge] (b2) edge[pos=0.8,"$h_{22}$"] (t2) ;
    \path[uedge] (b1) edge[pos=0.9,"$h_{31}$"] (t3) ;
    \path[uedge] (b2) edge[pos=0.5,"$h_{32}$"] (t3) ;
    \node at (0.5,-0.3) {Decoder $\mathbf{q}_j=[q_{j,1},\dots, q_{j,T}]$};
    \node at (0.5, 1.3) {Precoder $\mathbf{p}_i=[p_{i,1},\dots, p_{i,T}]$};
    \node at (-0.9,1) {Senders};
    \node at (-0.9,0) {Receivers};
  \end{scope}
\end{tikzpicture}        
		\caption{An example of the system model with $3$ senders and $2$ receivers. $T$ is the number of communication slots.}
		\label{fig:example}
	\end{figure}
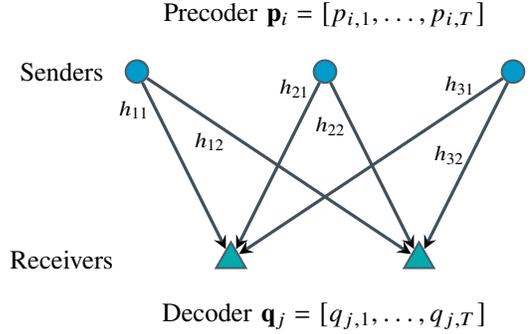

	Let $\mathcal{S}$ denote the set of senders with $|\mathcal{S}|=N_s$ and
	$\mathcal{R}$ denote the set of receivers with $|\mathcal{R}|=N_r$. The set of
	directed communication links is denoted by $\mathcal{E}$.\footnote{We consider a network where not all senders can communicate with all receivers. This may happen in a wireless network with spatially distributed nodes over a large geographical area, where some links with very large distances can be considered as inactive. In the special case where all senders can communicate with all receivers, we have $\mathcal{E}=\{(i,j),\forall i\in\mathcal{S}, j\in\mathcal{R}\}$.  }
	We use $\mathcal{N}_j=\{i\in\mathcal{S}|(i,j)\in\mathcal{E}\}$ to denote the set of senders that can communicate with receiver $j$. For each link $(i,j)\in\mathcal{E}$, the channel gain of the wireless link from sender $i$ to receiver $j$ is $h_{ij}\in\mathbb{C}$, which is known to the sender. We also assume additive White Gaussian noise (AWGN) with zero mean and variance $\sigma^2$ in all the channels. An example of the system model is illustrated in Fig.~\ref{fig:example}.
	
	We assume that each sender $i\in\mathcal{S}$ has a deterministic data sample $s_i\in\mathbb{C}$.\footnote{In general, the data samples can be both scalar or vector, with minor difference in the notations and analysis. In this paper, we only consider the scalar case to illustrate the concept of our proposed design. }
	The goal is to compute at each receiver $j\in\mathcal{R}$, the average of the data samples from its connected senders, i.e., 
	\begin{equation}
		\theta_j=\frac{1}{|\mathcal{N}_j|}\sum_{i\in\mathcal{N}_j} s_i,
	\end{equation}
	as close as possible. Here, $|\mathcal{N}_j|$ refers to the cardinality of the set $\mathcal{N}_j$.
	
	In a standard OtA system with multiple senders and one receiver, each sender precodes its data sample by multiplying it with a complex scalar to proactively compensate for the amplitude degradation and phase rotation of signals in wireless channels. The scaling factor is inversely proportional to the channel gain of the link between each sender and the common receiver, subject to the power constraint of the sender.
	
	\subsection{Multi-Slot Precoding and Decoding Design}
	With multiple receivers in the system, it is generally difficult to achieve simultaneous computations at different receivers within one slot, since the precoding factor of one sender cannot be adapted to multiple channels at the same time.
	One solution is to allow transmissions over multiple time slots and exploit the degrees of freedom in time.
	In each slot $t= \{1,\ldots,T\}$, sender $i$ transmits its precoded data $s_i p_{i,t}$, where $p_{i,t}\in\mathbb{C}$ is the corresponding precoding factor of sender $i$ in slot $t$. The transmitted signals must satisfy the total power constraint
	\begin{equation}
		|s_i|^2  \sum_{t=1}^{T}  |p_{i,t} |^2\leq P_{\max}.
	\end{equation}

	At the receiver side, in each slot $t$, receiver $j\in\mathcal{R}$ receives the following noisy superimposed signal
	\begin{equation}
		y_{j,t}=\sum_{i\in\mathcal{N}_j} s_i p_{i,t}  h_{ij}+n_{j,t},
	\end{equation}
	where $n_{j,t}\sim \mathcal{CN}(0, \sigma^2)$ is the AWGN noise at receiver $j$ in slot $t$. After receiving the signals over $T$ slots, receiver $j$ performs decoding by the following rule
	\begin{equation}
		\hat{\theta}_j=\sum_{t=1}^{T}q_{j,t}y_{j,t},
	\end{equation}
	where $q_{j,t}\in\mathbb{C}$ is the decoding factor of receiver $j$ in slot $t$.
	
	To obtain an unbiased estimate of $\theta_j$ for all $j\in\mathcal{R}$, the precoding and decoding factors must satisfy
	\begin{equation}
		\sum_{t=1}^{T} p_{i,t}  q_{j,t}=\frac{1}{|\mathcal{N}_j|  h_{ij}}, \forall (i,j)\in \mathcal{E}.
		\label{eq:unbiased}
	\end{equation}
	When the above condition is satisfied, the mean squared error (MSE) of the OtA computation averaged over all receivers is given by
	\begin{align}
		\textnormal{MSE}&=\frac{1}{N_r}\sum_{j=1}^{N_r}\mathbb{E}\left[|\hat{\theta}_j-\theta_j|^2\right]  \nonumber \\
		&=\frac{\sigma^2}{N_r} \sum_{j=1}^{N_r}\sum_{t=1}^{T}\lvert q_{j,t}\rvert^2.
	\end{align}

	\subsection{Similarities and Differences with Other Diversity Techniques}
	The proposed multi-slot communication model can be seen as one particular example of precoding over a vector space. From this perspective, our proposed design shares similarities with other diversity techniques such as precoding over multiple antennas or multiple sub-carriers. However, from the perspective of channel modeling, our proposed design is very different from the alternative ones. We assume that the channels remain constant within a certain duration $T$, which is smaller than the coherence time. As the result, the channel gain between each pair of sender and receiver is a constant scalar, based on which we obtain the unbiasedness condition in \eqref{eq:unbiased}.
		
	For the multi-antenna case, with the same dimension of the precoding and decoding design, we need $T$ transmitting antennas at each sender and $T$ receiving antennas at each receiver. The dimension of the channels becomes $(N_r\times T)\times (N_s\times T)$ instead of $N_r\times N_s$. Let $\mathbf{p}_i=[p_{i,1},\ldots,p_{i,T}]^{\top}$ represent the precoding vector of sender $i$ and $\mathbf{q}_j=[q_{j,1},\ldots,q_{j,T}]^{\top}$ represent the decoding vector of receiver $j$. Then, to obtain an unbiased estimate, the precoding and decoding vectors need to satisfy the following relation
	\begin{equation}
		\mathbf{p}_i^{\top}\mathbf{H}_{ij}\mathbf{q}_j=\frac{1}{|\mathcal{N}_j|}, \label{eq:condition}
	\end{equation}
	where $\mathbf{H}_{ij}$ is a $T\times T$ channel matrix between sender $i$ and receiver $j$, and every element in the matrix is non-zero. With our proposed design, we have $\mathbf{H}_{ij}=h_{ij}\cdot \mathbf{I}$ where $h_{ij}$ is a constant and $\mathbf{I}$ is a $T\times T$ identify matrix. 
	
	When $\mathbf{H}_{ij}$ is a scaled identity matrix, the feasible set (the set of $\mathbf{p}_i$ and $\mathbf{q}_j$ that satisfy the equality condition in \eqref{eq:condition}) is much larger. This is the main advantage of our multi-slot design as compared to the multi-antenna case. In practice, obtaining channel state information of $(N_r\times T)\times (N_s\times T)$ channels is also much more challenging than $N_r\times N_s$ channels.
	
	
	\section{Problem Formulation and Algorithm Design}
	We define an optimization problem that aims at minimizing the MSE of the OtA computation under the power constraints of the senders, for a given number of time slots $T$.
	The optimization variables are the precoding vectors $\mathbf{p}_i=[p_{i,1},\ldots,p_{i,T}]^{\top}$, $\forall i\in\mathcal{S}$ and the decoding vectors $\mathbf{q}_j=[q_{j,1},\ldots,q_{j,T}]^{\top}$, $\forall j\in\mathcal{R}$.
	The optimization problem is formulated as
	\begin{subequations}
		\label{eq:opt-1}
		\begin{align}
			\minimize~~&\sum_{j=1}^{N_r} \|\mathbf{q}_j\|^2\\
			\textrm{subject~to}~~& \mathbf{p}_i^\top  \mathbf{q}_j=w_{ij}, \forall (i,j)\in \mathcal{E}   \label{cons1}\\
			&\|\mathbf{p}_i\|^2\leq C_i, \forall i\in\mathcal{S} \label{cons2}.
		\end{align}
	\end{subequations}
	Here, $w_{ij}=\frac{1}{|\mathcal{N}_j|\cdot h_{ij}}$ for each link $(i,j)$ and $C_i=\frac{P_{\max}}{|s_i|^2}$ for each sender $i$. The constraint in \eqref{cons1} is the result of the unbiasedness condition in \eqref{eq:unbiased}, and \eqref{cons2} comes from the total power limit of each sender.

	This problem can be re-formulated as follows. We define $\mathbf{P}=[\mathbf{p}_1, \mathbf{p}_2,\ldots, \mathbf{p}_{N_s}]\in \mathbb{C}^{T\times N_s}$, $\mathbf{Q}=[\mathbf{q}_1, \mathbf{q}_2,\ldots, \mathbf{q}_{N_r}]\in \mathbb{C}^{T\times N_r}$, and $\mathbf{W}\in\mathbb{C}^{N_s\times N_r}$, where the $i$-th row and $j$-th column of $\mathbf{W}$ is
	\begin{equation}
		[\mathbf{W}]_{ij}=
		\left\lbrace
		\begin{array}{ccc}
			w_{ij}
			& \text{if}~(i,j)\in \mathcal{E}, \\
			0
			& \text{otherwise}.
			\label{eq:weight}
		\end{array} \right.
	\end{equation}
	Then the problem becomes
	\begin{subequations}
		\begin{align}
			\minimize\limits_{\mathbf{P}, \mathbf{Q}}~~&\|\mathbf{Q}\|_{F}^2\\
			\textrm{subject~to}~~& \mathbf{P}^\top \mathbf{Q}= \mathbf{W}  \label{eq:constraint1}\\
			&\|\mathbf{p}_i\|^2\leq C_i, i=1,\ldots, N_s.
			\label{eq:opt-2}
		\end{align}
	\end{subequations}
	Clearly, the latter is an instance of matrix factorization
	problems. However, in this form it is not much tractable due to the
	highly non-convex hard constraint~\eqref{eq:constraint1}. Instead, we consider the penalized version
	\begin{subequations}
		\begin{align}
			\minimize\limits_{\mathbf{P}, \mathbf{Q}}~~&\|\mathbf{P}^\top \mathbf{Q}- \mathbf{W}\|_F^2+\lambda \|\mathbf{Q}\|_{F}^2\\
			\textrm{subject~to}~~
			&\|\mathbf{p}_i\|^2\leq C_i, i=1,\ldots, N_s,
			\label{eq:opt-3}
		\end{align}
	\end{subequations}
	where $\lambda > 0$ is a regularization parameter.\footnote{The solution to this penalized problem cannot guarantee that the unbiasedness condition in \eqref{eq:constraint1} is satisfied. However, in simulation results we will show that this will not significant affect the MSE.} Now we have convex constraints, but nonconvex objective function. However,
	this type of problems is known to be tractable by standard
	first-order methods \cite{gillis2020nonnegative}.

	As we have simple convex constraints and differentiable objective, we could
	apply the projected gradient method. Note that the orthogonal projection onto this
	constraint is defined as
	$\Pi(\mathbf{P}) = \left[\Pi_{B(\sqrt{C_i})}(\mathbf{p_i})\right]_{i\in\mathcal{S}}$,
	which is the projection of $i$-th column of $\mathbf{P}$ onto the ball $B(\sqrt{C_i})$.
	However, the projected gradient method requires the step size as an
	input parameter. In our case, it is non-trivial since the 
	gradient of the objective function is not Lipschitz continuous. For this reason,
	we apply the adaptive version of the projected gradient descent method from~\cite{adgd}.\footnote{Note that the method we use is theoretically sound only in the convex case. The problem we consider is a particular case of the matrix factorization problem, and although it is nonconvex, first-order optimization methods can solve such kind of problems successfully.}
	 The details of the developed algorithm are described in Algorithm~\ref{alg:main-1}.
	
	
	\begin{algorithm}[t]
		\caption{: Adaptive projected gradient descent for $\min_{\mathbf{X}\in \mathcal{X}} f(\mathbf{X})$}
		\label{alg:main-1}
		\begin{algorithmic}[1]
			\STATE \textbf{Input:} $\mathbf{X}^{-1}, \mathbf{X}^0 \in \mathbb{R}^d$,
			$\alpha_{-1}=\alpha_0>0$
			\FOR{$k = 1,2,\dots$ (until stopping criteria)}
			\STATE $\alpha_k = \min\Bigl\{\sqrt{1+\frac{\alpha_{k-1}}{\alpha_{k-2}}} \alpha_{k-1},
			\frac{\|\mathbf{X}^{k}-\mathbf{X}^{k-1}\|}{2\|\nabla f(\mathbf{X}^{k})-\nabla f(\mathbf{X}^{k-1})\|}\Bigr\}$
			\STATE $\mathbf{X}^{k+1} = \Pi_{\mathcal{X}}\left(\mathbf{X}^k - \alpha_k \nabla f(\mathbf{X}^k)\right)$
			\ENDFOR
			\STATEx \textbf{Implementation details:}\\
			$\mathbf{X} = [\mathbf{P}, \mathbf{Q}]$\\
			$f(\mathbf{X}) = \|\mathbf{P}^\top \mathbf{Q} - \mathbf{W}\|^2_F + \lambda \|\mathbf{Q}\|^2_F$\\
			$\nabla f(\mathbf{X}) = 2[\overline{\mathbf{Q}} (\mathbf{P}^\top \mathbf{Q}-\mathbf{W})^\top,  \overline {\mathbf{P}}(\mathbf{P}^\top \mathbf{Q}-\mathbf{W}) + \lambda \mathbf{Q}]$\\
			$\Pi_{\mathcal{X}}(\mathbf{X}) = [\Pi(\mathbf{P}), \mathbf{Q}]$
		\end{algorithmic}
	\end{algorithm}

	
	
	\section{Simulation Results}
	
	To verify the performance of our proposed method, we consider the following simulation setting. The network consists of $N_s = 50$ senders and $N_r = 30$ receivers. Each sender is associated with $20$ receivers (randomly selected from the set of receivers).  
	The data samples of the senders are randomly drawn from $\mathcal{CN}(0,1)$ distribution and are fixed in all simulations. The channel gain of each communication link follows $\mathcal{CN}(0,1)$ distribution.
	The signal-to-noise ratio (SNR) is $P_{\max}/\sigma^2 = \{1,10,100\}$. The regularization parameter is chosen as $\lambda=0.1$. The results are averaged over $100$ channel realizations.\footnote{Our simulation code is available at \href{https://github.com/ymalitsky/space-time-ota}{https://github.com/ymalitsky/space-time-ota}.}

	As mentioned in the Introduction, a baseline approach for achieving OtA computation with multiple receivers is to divide the network into several parallel sub-networks by assigning one time slot to each receiver. In every slot (assigned to one specific receiver), we follow the procedure of the standard OtA computation. In this case, we need $T=N_r$ time slots in total. 
	
\begin{figure}[t!]
		\centering
		\begin{subfigure}[b]{0.5\textwidth}
			\includegraphics[width=0.8\textwidth]{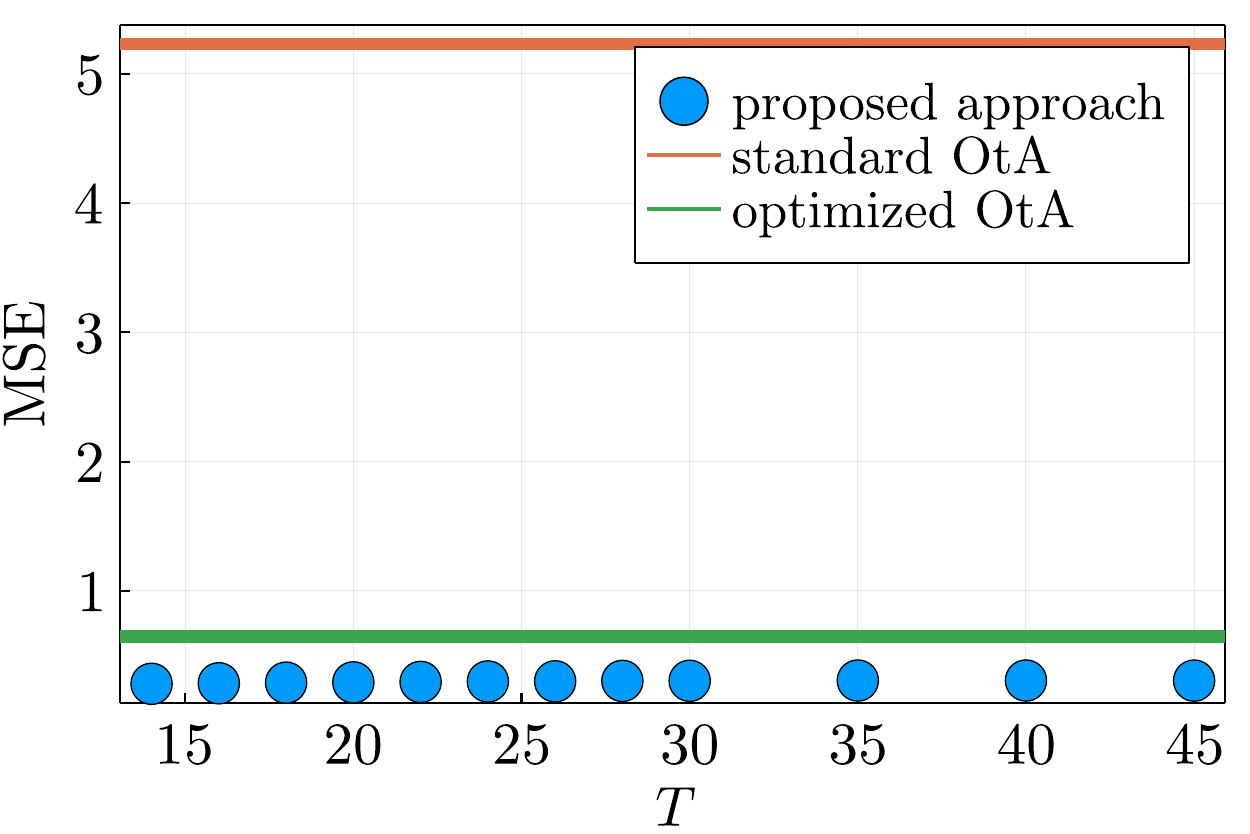}
		\end{subfigure}
		\hfill
		
		\begin{subfigure}[b]{0.5\textwidth}
			\includegraphics[width=0.8\textwidth]{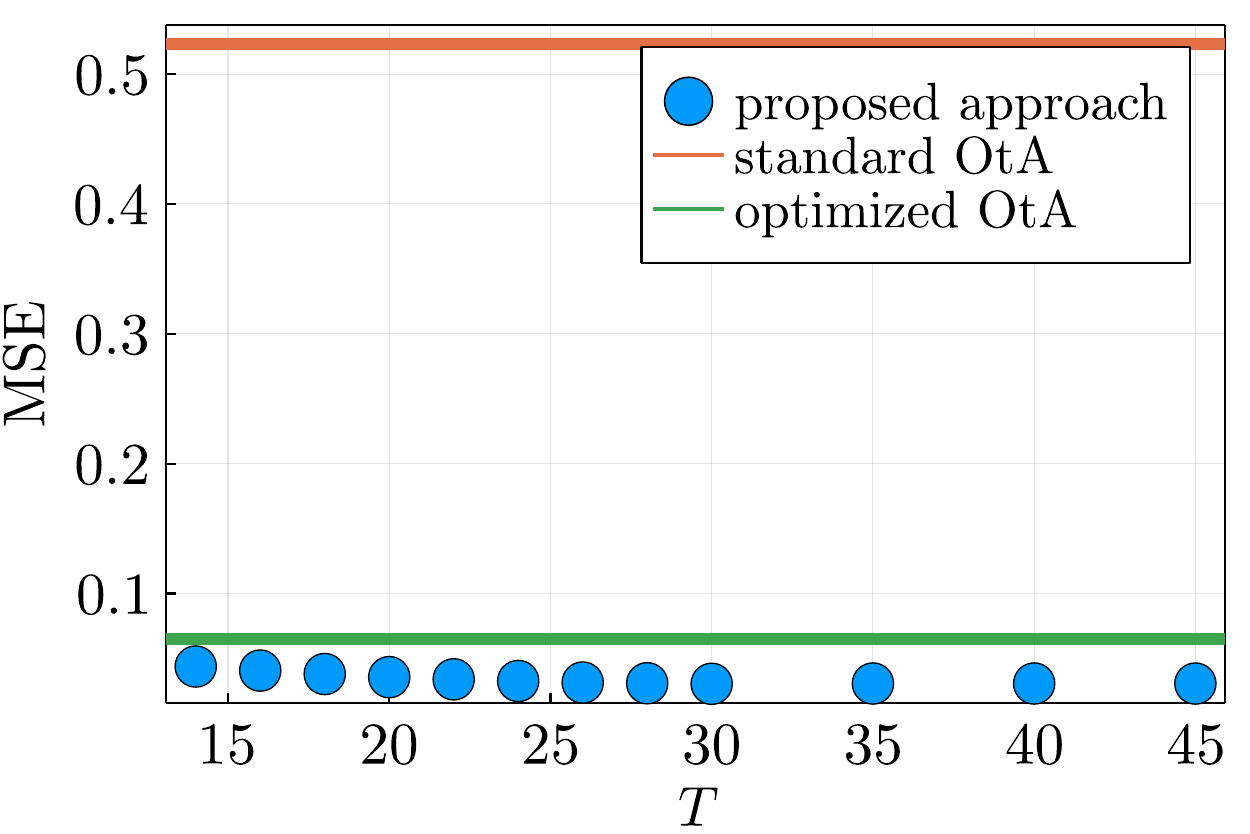}
		\end{subfigure}
		\hfill
		
		\begin{subfigure}[b]{0.5\textwidth}
			\includegraphics[width=0.8\textwidth]{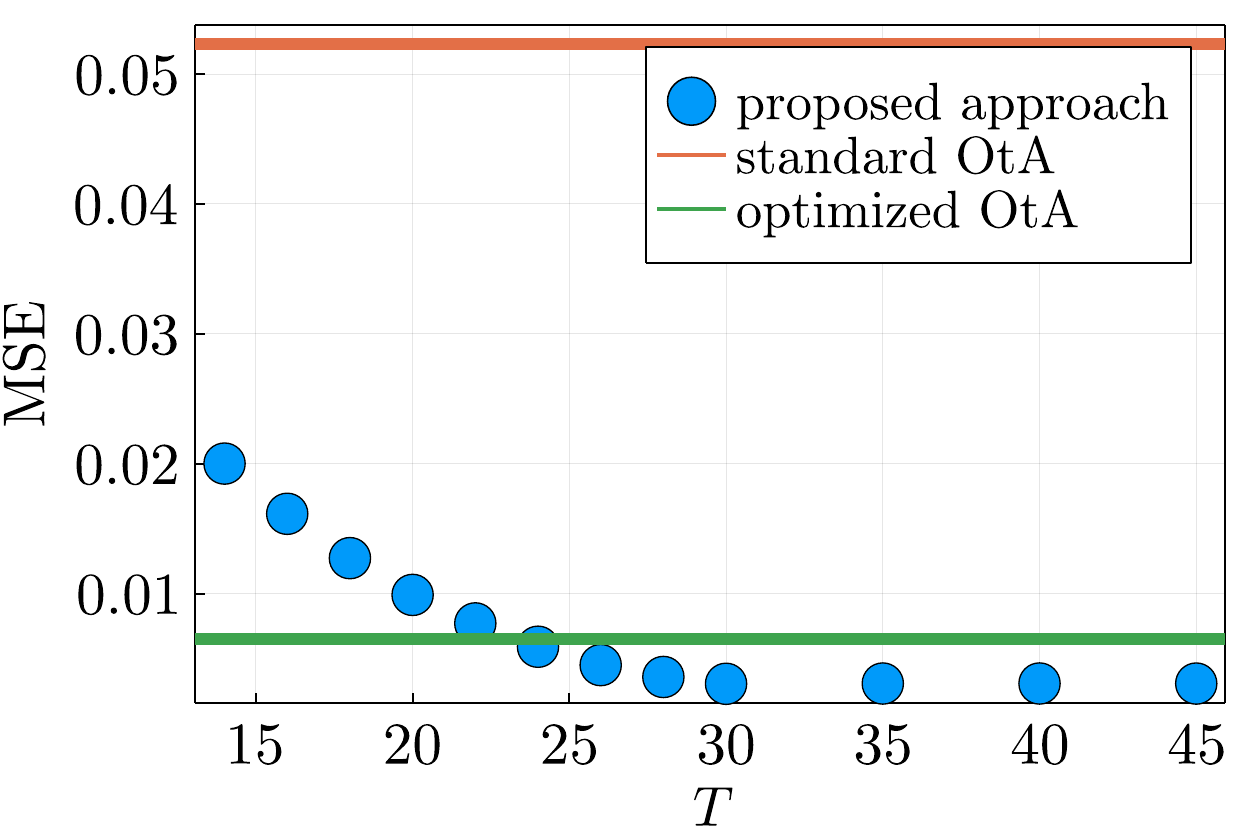}
		\end{subfigure}
		\caption{MSE of our proposed multi-slot design for different values of  $T$ (number of slots). The results of baseline OtA approaches are obtained with $T=30$. } 
		\label{fig:1}
	\end{figure}

	For performance comparison, we have implemented two versions of the baseline OtA approach:
	\begin{itemize}
		\item The power budget of each sender $P_{\max}$ is equally spread over its connected receivers, i.e., each sender has power constraint of $P_{\max}/20$ within each sub-network. The results are marked as ``standard OtA''.
		\item The power budget of each sender $i$ allocated for the computation of receiver $j$ is obtained by minimizing the average MSE subject to the total power constraint of each sender. The problem is defined as
		\begin{subequations}
			\begin{align}
				\minimize_{p_{ij}>0}~~&\frac{1}{N_r} \sum_{j=1}^{N_r}\max_{i\in \mathcal{N}_j}\left\{ \frac{\sigma^2 w_{ij}^2|s_i|^2}{p_{ij} |h_{ij}|^2}\right\}\\
				\text{subject
					to}~~&\sum_{j=1}^{N_r}p_{ij}\leq P_{\max},
			\end{align}
		\end{subequations}
		where $p_{i,j},\forall (i,j)$ represents the power constraint of sender $i$ for the computation of receiver $j$. The results are marked as ``optimized OtA''.
	\end{itemize} 
	The simulation results are presented in Fig.~\ref{fig:1}. Note that in ``standard OtA'' and ``optimized OtA'' cases, we have only one value at $T=N_r=30$, but the results are illustrated as two straight lines for comparison purpose.

	\begin{figure}[t!]
		\centering
		\begin{subfigure}[b]{0.5\textwidth}
			\includegraphics[width=0.8\textwidth]{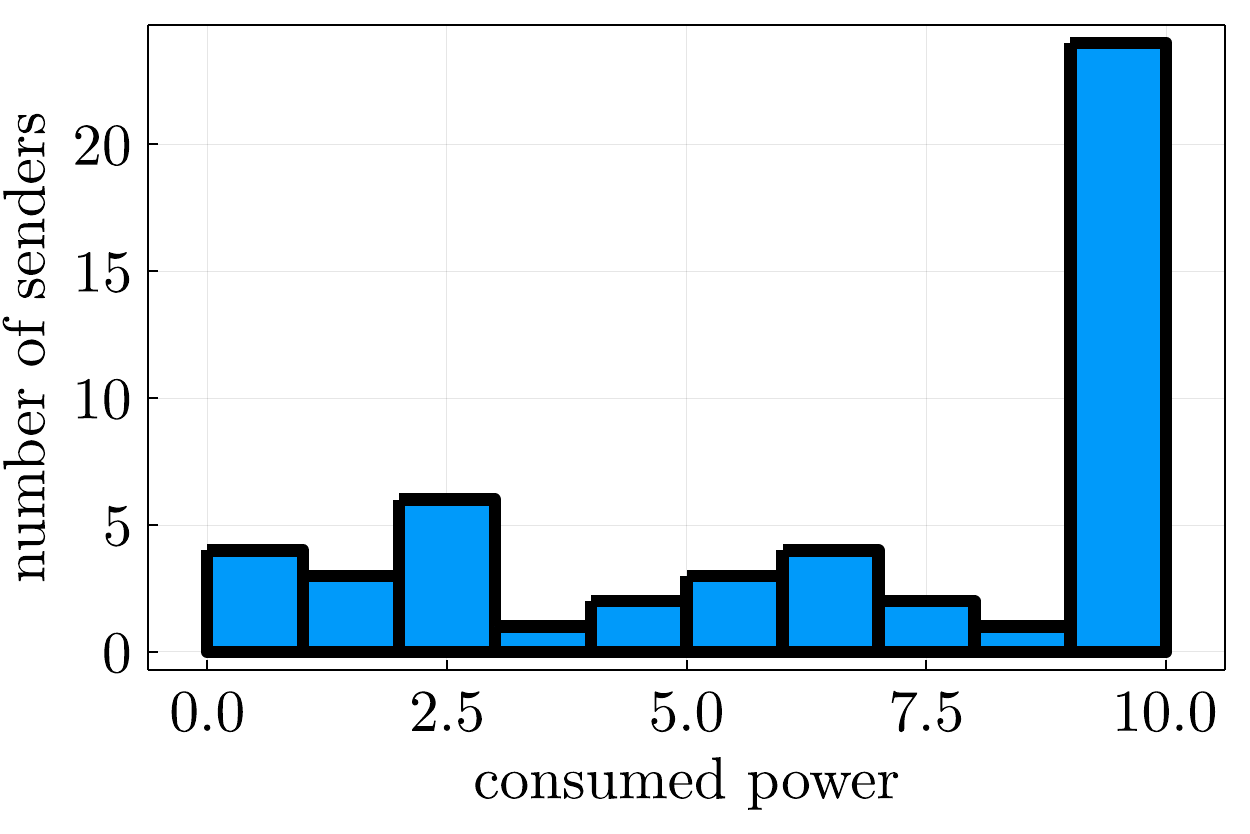}
		\end{subfigure}
		\hfill
		
		\begin{subfigure}[b]{0.5\textwidth}
			\includegraphics[width=0.8\textwidth]{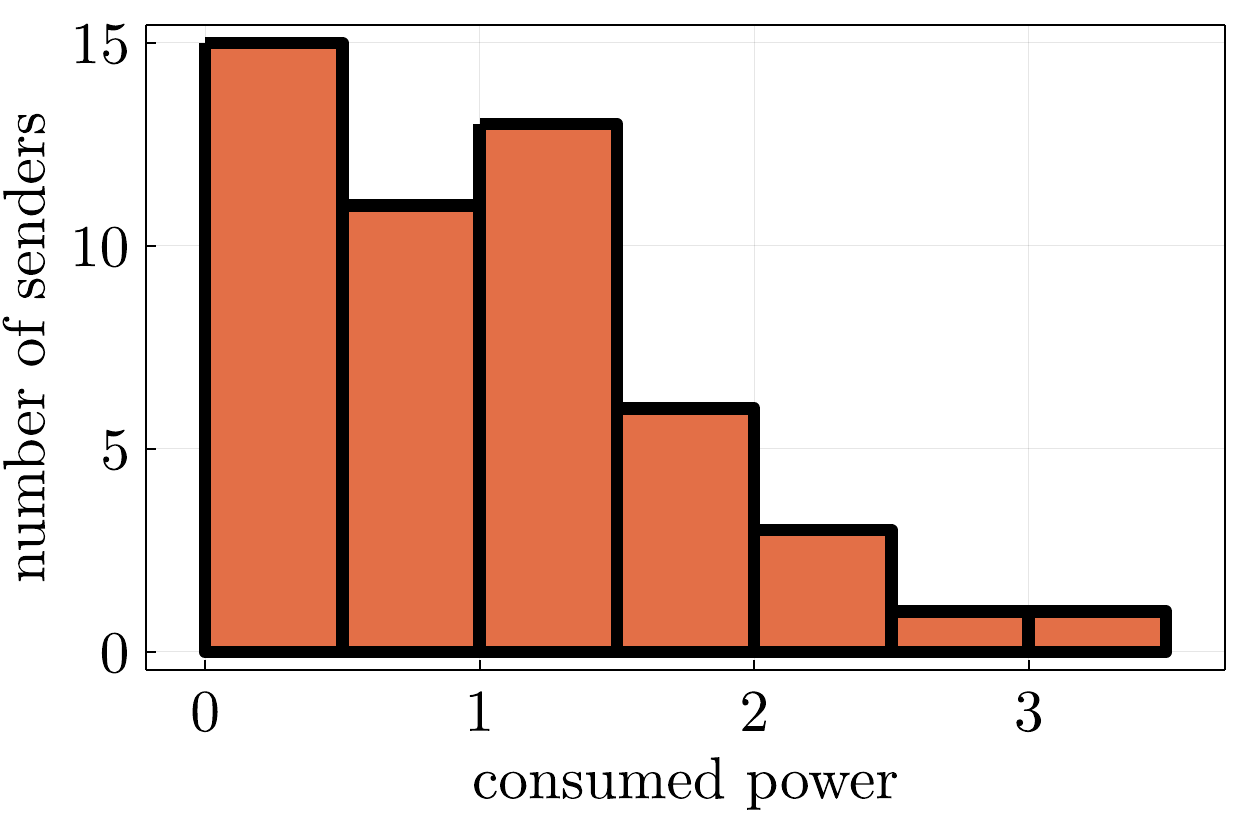}
		\end{subfigure}
		\begin{subfigure}[b]{0.5\textwidth}
			\includegraphics[width=0.8\textwidth]{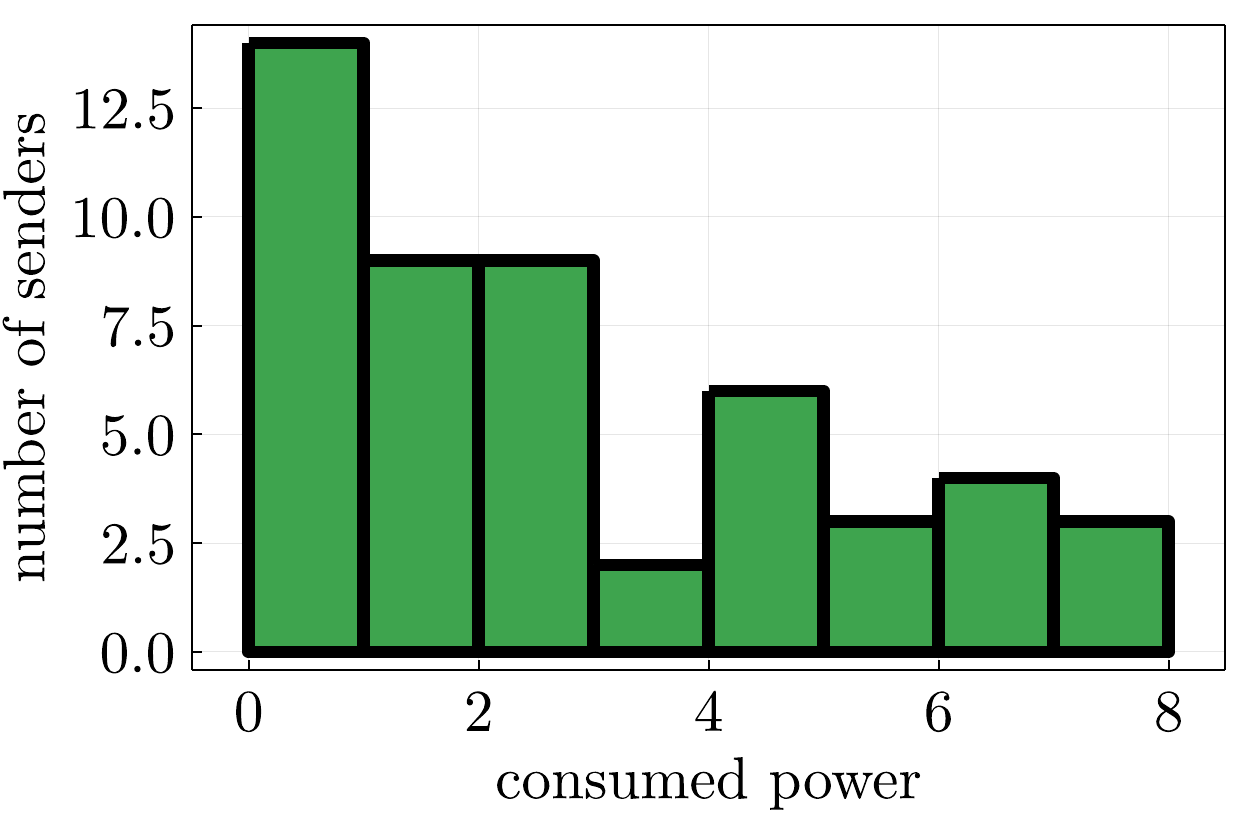}
		\end{subfigure} 
		\caption{Power consumption: proposed (above), standard OtA (middle), and
			optimized OtA (bottom)
			approaches.}
		\label{fig:2}
	\end{figure}   
	
	We see that with our multi-slot joint precoding and decoding design, we can achieve much lower MSE than the standard OtA design for all SNR values. With $T=N_r=30$, our approach gives only $5.8\%$ of the estimation error as compared to the standard OtA case with equal power budget allocation. Another observation is that the performance of the standard parallel OtA design can be improved by optimizing the power budget allocation over different receivers. However, our proposed design still shows clear advantage as compared to the optimized OtA case. This also means that to obtain a good level of estimation error, our proposed design will require much less communication resources (e.g., time slots) than separating different receivers over time. 
	
	To better understand this significant performance gain that we observe, in Fig.~\ref{fig:2}, we plot the histogram of the actual power consumption of the senders for our proposed design and the two baseline OtA approaches. Clearly, our proposed design allows many senders to reach their full power limit, while with the baseline OtA approaches, a large fraction of the available power is unconsumed. One reason is that the common amplitude scaling factor $\eta$ in \eqref{eq:eta} creates the bottleneck effect where only one sender can actually reach the power limit. Optimizing the power budget allocation of each sender for different receivers (sub-networks) can alleviate this bottleneck problem, but not entirely solve it. Most importantly, our design is valid for any value of $T$, while the baseline approaches are only meaningful for $T=N_r$. 
	
	\section{Conclusions}
	In this work, we proposed a multi-slot joint precoding and decoding design for Over-the-Air (OtA) computation in multiple-receiver scenarios. As compared to the baseline approach with orthogonal time division among different receivers, our proposed multi-slot design offers significant reductions in communication time while achieving a lower estimation error. The design can be adapted for achieving OtA computation in complex networks with non-trivial topology, e.g., when a node can serve as both a sender and a receiver with full-duplex capability. The basic concept can also be extended to other scenarios with imperfect channel state information, and multiple antennas at the sender and receiver sides. The scalability of this design in large-scale wireless networks is also an important aspect for future research.


\end{document}